\documentclass[pdflatex,sn-nature]{sn-jnl}% Style for 

\usepackage{graphicx}%
\usepackage{multirow}%
\usepackage{amsmath,amssymb,amsfonts}%
\usepackage{amsthm}%
\usepackage{mathrsfs}%
\usepackage[title]{appendix}%
\usepackage{xcolor}%
\usepackage{textcomp}%
\usepackage{manyfoot}%
\usepackage{booktabs}%
\usepackage{algorithm}%
\usepackage{algorithmicx}%
\usepackage{algpseudocode}%
\usepackage{listings}%
\usepackage{subfiles}

\usepackage{xr}
\usepackage{xcolor}

\newif\ifshowrevisions
\showrevisionsfalse  % <-- Change to \showrevisionsfalse to show revisions in black

\ifshowrevisions
  \newcommand{\rev}[1]{\textcolor{red}{#1}}
\else
  \newcommand{\rev}[1]{#1}
\fi
\begin{document}

\title[Article Title]{Solving Boolean satisfiability problems with resistive content addressable memories}

\author*[1]{\fnm{Giacomo}\sur{Pedretti}}\email{giacomo.pedretti@hpe.com}
\author[2]{\fnm{Fabian} \sur{B{\"o}hm}}
\author[3,4]{\fnm{Tinish} \sur{Bhattacharya}}
\author[5]{\fnm{Arne} \sur{Heittman}}
\author[6]{\fnm{Xiangyi} \sur{Zhang}}
\author[5]{\fnm{Mohammad} \sur{Hizzani}}
\author[3]{\fnm{George} \sur{Hutchinson}}
\author[3]{\fnm{Dongseok} \sur{Kwon}}
\author[1]{\fnm{John}\sur{Moon}}
\author[6]{\fnm{Elisabetta} \sur{Valiante}}
\author[6]{\fnm{Ignacio} \sur{Rozada}}
\author[1]{\fnm{Catherine E.}\sur{Graves}}
\author[1]{\fnm{Jim}\sur{Ignowski}}
\author[4]{\fnm{Masoud} \sur{Mohseni}}
\author[5]{\fnm{John Paul} \sur{Strachan}}
\author[3]{\fnm{Dmitri} \sur{Strukov}}
\author[4]{\fnm{Ray} \sur{Beausoleil}}
\author*[2]{\fnm{Thomas} \sur{Van Vaerenbergh}}\email{thomas.van-vaerenbergh@hpe.com}

\affil*[1]{\orgdiv{Artificial Intelligence Research Lab (AIRL)}, \orgname{Hewlett Packard Labs}, \orgaddress{\city{Milpitas}, \state{CA}, \country{United States}}}

\affil[2]{\orgdiv{Large Scale Integrated Photonics (LSIP)}, \orgname{Hewlett Packard Labs}, \orgaddress{ \city{Brussels}, \country{Belgium}}}

\affil[3]{\orgdiv{University of California Santa Barbara}, \orgname{(UCSB)}, \orgaddress{\city{Santa Barbara}, \state{California}, \country{United States}}}

\affil[4]{\orgdiv{Large Scale Integrated Photonics (LSIP)}, \orgname{Hewlett Packard Labs}, \orgaddress{\city{Milpitas}, \state{CA}, \country{United States}}}

\affil[5]{\orgdiv{Institute for Neuromorphic Compute Nodes (PGI-14), Peter Grunberg Institute}, \orgname{Forschungszentrum Juelich GmbH}, \orgaddress{\city{Juelich}, \country{Germany}}}

\affil[6]{\orgname{1QB Information Technologies (1QBit)}, \orgaddress{\city{Vancouver}, \state{British Columbia}, \country{Canada}}}

\abstract{Solving optimization problems is a highly demanding workload requiring high-performance computing systems. Optimization solvers are usually difficult to parallelize in conventional digital architectures, particularly when stochastic decisions are involved. Recently, analog computing architectures for accelerating stochastic optimization solvers have been presented, but they were limited to academic problems in quadratic polynomial format. Here we present KLIMA, a $k-$\textbf{L}ocal \textbf{I}n-\textbf{M}emory \textbf{A}ccelerator with resistive Content Addressable Memories (CAMs) and Dot-Product Engines (DPEs) to accelerate the solution of high-order industry-relevant optimization problems, in particular Boolean Satisfiability. By co-designing the optimization heuristics and circuit architecture we improve the speed and energy to solution up to $182\times$ compared to the digital state of the art.}


\maketitle
\newpage
\section{Introduction}\label{sec1}

\begin{figure}
    \centering
    \includegraphics[width=0.99\linewidth]{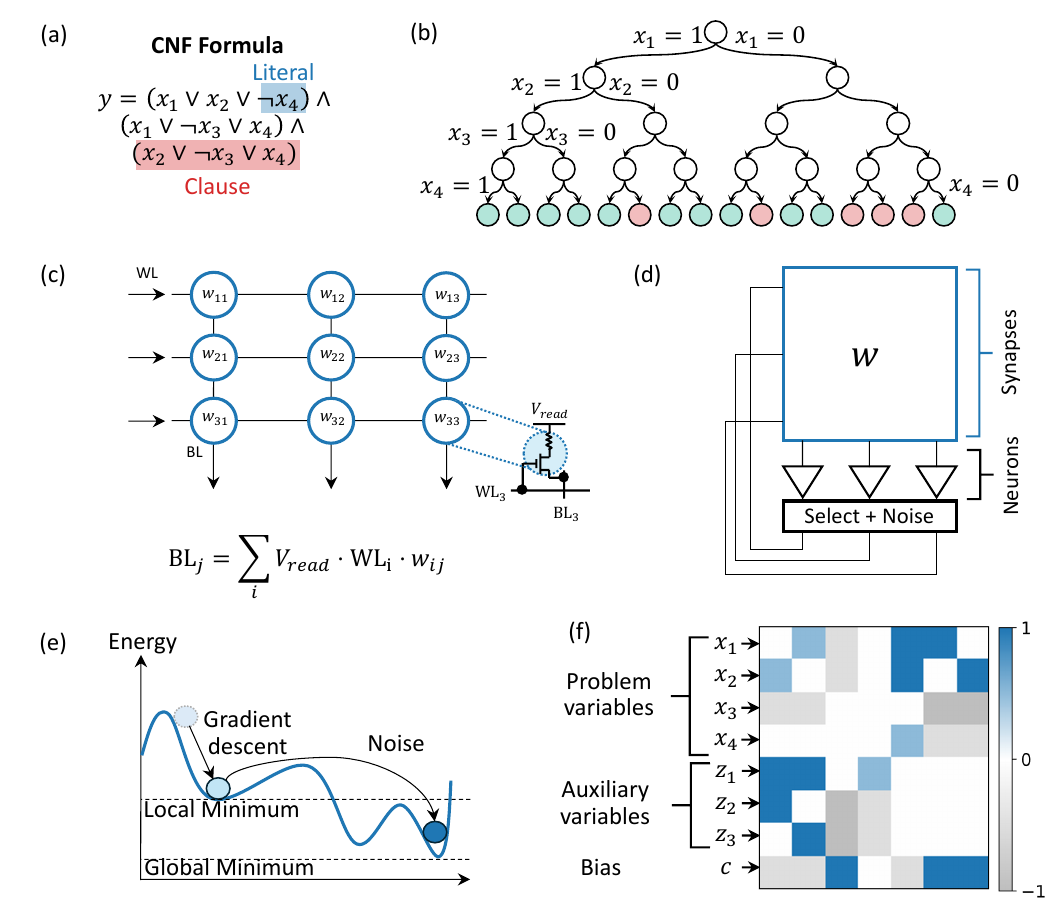}
    \caption{(a) Conjunctive Normal Form (CNF) of a Boolean Satisfiability problem. (b) Solution tree of the formula in (a). Green leaves are satisfied solutions while red leaves are unsatisfied solutions. (c) Crossbar array of RRAM implementing a Dot Product Engine (DPE). (d) Hopfield Neural Network (HNN) for optimization using a DPE. (e) Example of energy landscape navigation with an HNN. (f) Mapping of the problem in (a) into a QUBO formulation for being solved with an HNN.}
    \label{fig:intro}
\end{figure}
Boolean Satisfiability (SAT) problems are the backbone of several industry tasks.
SAT consists of finding a set of variables that satisfy a number of Boolean clauses~\cite{marques2008practical}.
Many problems are classically mapped to SAT, such as software and hardware verification~\cite{burch1992symbolic}, planning~\cite{rintanen2011planning} and scheduling~\cite{horbach2012using}.
More recently, additional interest in solving SAT problems has arisen because of the increased focus on reasoning in AI models, which are being equipped with conventional symbolic techniques to increase performance and reliability, a technique often referred to as neurosymbolic AI \cite{trinh2024solving,yang2024fine}.

SAT problems are usually represented in their Conjunctive Normal Forms (CNF).
Each clause is a disjunction (logical OR) of literals (i.e., normal or complementary Boolean variables), that becomes satisfied if at least one literal is true.
Multiple clauses are in conjunction (logical AND), leading to satisfaction of the formula if all the clauses are satisfied.
Fig. \ref{fig:intro}a shows an example of a CNF of a $k-$SAT problem with $V=4$ variables, $C=3$ clauses and order, namely the maximum number of variables in each clause, $k=3$.

Solving a SAT problem involves finding the set of variables $x_i$ that satisfies the CNF formula. A SAT problem may have none, one, or multiple solutions. Fig. \ref{fig:intro}b shows the solution tree for the problem in Fig.\ref{fig:intro}a. At each node, a variable is assigned either 0 or 1. Leaves, or variables set, corresponding to a satisfied problem are represented in green, while unsatisfied sets are represented in red. The solver algorithm decides how to assign each variable. Solvers can be divided into two main classes: exact solvers and Stochastic Local Search (SLS) solvers~\cite{hoos2000local}. Modern exact solvers rely on backtracking techniques~\cite{davis1962machine}, in which variables are assigned individually until a clause becomes unsatisfied, and the last assigned variable is reverted. Backtracking solvers are complete, i.e., if a problem is solvable it is guaranteed to find the solution; if it is not solvable, it can be demonstrated by showing which clauses conflict. However, exact solvers can be slow, especially where the search space is very large, or with very high order $k$, making exhaustive exploration impractical. 
SLS solvers start from a random assignment, updating one or more variables either randomly or deterministically based on a metric, for example, a gradient computation. Such metrics suggest the most promising variable(s) to improve the solution and the solver essentially optimizes the problem by minimizing the number of violated clauses until reaching a certain minimum. 
SLS solvers can be faster in converging in particular for random or hard-to-structure SAT instances, but they do not guarantee convergence to an exact solution. Recently, hybrid exact-SLS solvers have been proposed and are considered state-of-the-art~\cite{cai2021deep}. However, their stochastic and serial nature makes SLS solvers complicated to run and parallelize on conventional von Neumann architectures.

Recently, several analog accelerators for solving constraint satisfaction problems based on Ising Machines (IM)~\cite{mohseni2022ising} or Hopfield Neural Networks (HNNs)~\cite{hopfield1985neural} architecture with different technologies such as in-memory computing~\cite{cai2020power,mahmoodi2019versatile}, probabilistic bits~\cite{aadit2022massively}, and photonics~\cite{bohm2019poor} have been proposed.
In-memory computing~\cite{ielmini2018memory} is considered a promising and efficient paradigm, in which data is processed within the memory limiting the data transfer between the latter and the processing units.
In particular, emerging non-volatile memories, such as Resistive Random Access Memories (RRAM, or memristors)~\cite{ielmini2016resistive}, Phase Change Memories (PCM)~\cite{burr2016recent}, and others, are compact, energy-efficient, and can be programmed in an analog fashion, making them an ideal candidate for in-memory computing implementation.
By organizing them in crossbar arrays, for example, RRAMs can be used to encode a matrix, and Vector-Matrix-Multiplications (VMMs) can be performed in the analog domain~\cite{ielmini2020device}.
\rev{HNNs are characterized by VMMs between an input vector (or neuron state) and a matrix storing the coupling terms between states, making them computationally intensive in traditional digital architectures.
The neuron states are binary, making crossbar array implementations of HNNs particularly attractive given that not only the VMMs can be efficiently accelerated but analog to digital converters are not required, and simple comparators can be used as neuron circuits}~\cite{cai2020power}. 
%Analog optimization accelerators can exploit such RRAM arrays for massively parallel computation of gradients of an optimization problem's cost function~\cite{cai2020power}. 
The cost function is formulated as a polynomial energy function and gradient descent methods can be implemented in hardware to find optimal solutions. 
Because of their high parallelism and ability to rapidly converge, such analog accelerators have also been considered for solving SAT problems.
Conventionally, these computing primitives can implement only quadratic interactions between variables, leading to significant overhead for solving SAT problems of order $k>2$~\cite{hizzani2024memristor}.
SAT problems with $k=2$, or 2-SAT, are not computationally intensive, belonging to the \textit{P} class of computational complexity in which the computational resources scale polynomially with the problem size, whereas $k$-SAT problems with $k>2$ belong to the \textit{NP} class of problems, where computational resources scale exponentially.
Moreover, the heuristics that HNNs/IMs implement are not optimized specifically for SAT and compared to state-of-the-art SLS for SAT, require significantly more iterations~\cite{bhattacharya2024computing}.

In this paper, we build on previous work where we first demonstrated the concept of solving SAT in its native formulation, using Content Addressable Memories (CAMs)~\cite{pedretti2022differentiable,pedretti2023zeroth,bhattacharya2024computing} for computing interactions between variables.
We perform hardware-software co-design of the heuristics and the required custom circuit blocks, such as for noise generation, highlighting the trade-off and optimizing for improved energy efficiency. 
We extensively benchmark the proposed hardware against state-of-the-art digital counterparts demonstrating up to 182$\times$ improved energy efficiency in solving random uniform 3-SAT and 4-SAT problems.

\section{Results}\label{sec:result}

\subsection{Inefficiencies in solving SAT with conventional HNNs}\label{sec:hnns}

Fig. \ref{fig:intro}c shows a crossbar array RRAMs for performing a VMM. 
Each RRAM cell represents a weight, or matrix parameter, $w_{ij}$.
We consider a 1-transistor-1-resistor (1T1R) structure, with the transistor acting as a selector, to increase the robustness to non-idealities~\cite{ielmini2020device}.
Binary inputs are applied to the 1T1R Word Lines (WLs) and the resulting current, accumulated on the Bit Lines (BLs), is the product of the input vector and parameters matrix.
Crossbar arrays can be used as building blocks for HNNs~\cite{cai2020power} and Ising Machines~\cite{chiang2024reaim}.
Fig. \ref{fig:intro}d shows a high-level circuit schematic of an HNN.
HNNs are energy-based recurrent associative memories that retrieve a stored pattern in an attractor basin, or energy minimum.
This recall ability can navigate rugged energy landscapes and optimize a problem by opportunely mapping it in the coupling terms between neurons~\cite{hopfield1985neural}.
At each iteration, a VMM operation is performed between the neuron state vector $x$ and the coupling matrix $w$ resulting in weighted accumulation as neuron input 
\begin{equation}
    y_j = \sum_{i}x_iw_{ij}.
\end{equation}
The neuron activation processes the input $y$ by comparing it with a threshold $\Theta_j$ leading to
\begin{equation}
    x_j = \begin{cases}
1 &\text{if $y_j\geq \Theta_j$}\\
0 &\text{otherwise}
\end{cases}
\end{equation}
The HNNs state is demonstrated to minimize the following energy function
\begin{equation}
    E = -\frac{1}{2}\sum_i\sum_jx_ix_jw_{ij} + \sum_i\Theta_ix_i
 . \end{equation}\label{eq:qubo}
However, during the gradient descent, it is possible to encounter a local minimum, thus noise is usually added to $y$, e.g. using Simulated Annealing~\cite{kirkpatrick1983optimization,cai2020power} heuristics, and optimize the given problem. 
Fig. \ref{fig:intro}e conceptually illustrates an example of energy landscape navigation, with gradient descent resulting in the solution stuck in a local minimum and how noise can be used to escape it.
In-memory accelerators of HNNs have been successfully demonstrated, mainly for the optimization of native quadratic unconstrained problems such as Max-Cut~\cite{cai2020power,mahmoodi2019versatile}.

Solving $k-$SAT poses a separate computational challenge.
Note that Eq. \ref{eq:qubo} has two terms, namely a quadratic term in which states from two neurons are multiplied, and a linear term, a mapping often referred to as Quadratic Unconstrained Binary Optimization (QUBO).
Thus for representing problems of order $k>2$, such as the ($k=3$)-SAT formula of Fig. \ref{fig:intro}a, as a QUBO problem, auxiliary variables have to be added~\cite{lucas2014ising}, as shown in Fig. \ref{fig:intro}f.
Adding auxiliary variables introduces several overheads.
First, the resulting problem has more variables, increasing the size of the search space and thereby making it more difficult to solve~\cite{valiante2021computational}.
Second, the circuit implementation would require more area and power to operate~\cite{hizzani2024memristor}.
Finally, mapping the auxiliary variables modifies the original energy landscape, which might result in novel local minima and saddle points, potentially making the problem even harder to solve~\cite{dobrynin2024energy}.

An improved HNN heuristic and circuit implementation that can map higher-order interactions was recently demonstrated~\cite{hizzani2024memristor,sharma2023augmenting,bybee2023efficient}, resulting in an HNN able to natively solve Polynomial Unconstrained Binary Optimization (PUBO) problems.
While leading to significant improvements compared to second-order HNNs~\cite{hizzani2024memristor} (HNN-Q), the polynomial heuristic (HNN-P) is still not specifically designed for SAT.
Moreover, such circuits~\cite{hizzani2024memristor,sharma2023augmenting} are specifically designed for a given order, e.g. $k\leq3$, by using hardwired interaction logic, and designing circuits for $k\gg3$ can lead to significant overhead.
Thus, they are less suitable for mapping arbitrary-order SAT problems and implementing competitive SAT-solving heuristics.

\subsection{$k$-Local Interactions In-Memory Accelerator (KLIMA)}\label{sec:klima}

\begin{figure}
    \centering
    \includegraphics[width=0.99\linewidth]{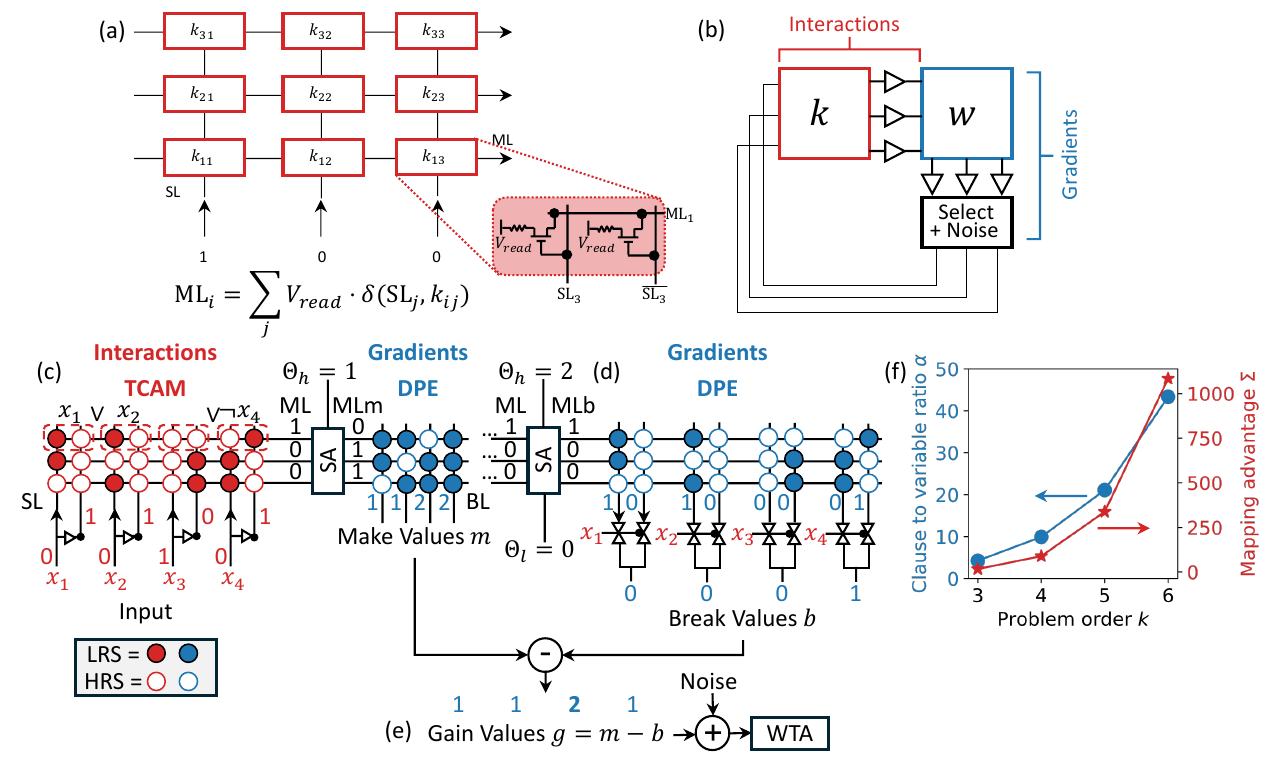}
    \caption{(a) Content Addressable Memory (CAM) realized with a 2T2R RRAM structure, with the output current on the ML representing the Hamming distance of the input and the stored word. (b) Proposed $k-$Local Interactions In-Memory Accelerator (KLIMA) using TCAM for computing interactions and DPE for computing gradients. (c) Example of computing the make value $m$ using KLIMA on the formula in Fig. \ref{fig:intro}. First, the interactions between variables are evaluated using the TCAM by computing \text{ML} which is then compared with a sense amplifier (SA) to threshold $\text{MLm}=\text{ML}<\Theta_h$ in order to evaluate the location of violated clauses. Then $\text{MLm}$ is used to compute $m$ with a dot product operation. (d) Computing the break value $b$ with KLIMA. The \text{ML} interactions are compared with two thresholds to evaluate which clauses are satisfied by only one literal, namely $\text{MLb}=(\Theta_l<\text{ML}<\Theta_h)$. $\text{MLb}$ is then used to compute the break count, by first accumulating the contribution of each member variable in a clause and comparing it with the current input $x$. (e) Computing the gain with KLIMA as $g=m-b$. (f) Mapping advantage of KLIMA compared to HNN in terms number of coupling terms $\Sigma$ as a function of order $k$ for random k-SAT problems at the computational phase transition.}
    \label{fig:klima}
\end{figure}

Recently, we presented a generalized HNN/IM and a Content Addressable Memory (CAM) able to map and accelerate higher-order heuristics such as WalkSAT~\cite{pedretti2023zeroth,bhattacharya2024computing,bhattacharya2024unified}.
CAMs are often referred to as complementary to Random Access Memories (RAMs).
In RAMs, an address is given as input and the data stored at that address is retrieved, while in CAMs data is given as input with its location, or address, retrieved as output.
SRAM-based CAMs are bulky and power-hungry, thus many RRAM-based CAM have been proposed with diverse circuit architectures~\cite{graves2022memory}.
Here, we focus on CAMs implemented with a RRAM -crossbar array~\cite{chen2015efficient} but the discussion can be extended to other circuit topologies and technologies as well.
Fig. \ref{fig:klima}a shows a crossbar array implementing a CAM.
Each CAM entry is mapped into a 2T2R cell, with the two RRAM complementary programmed into a High Resistance State (HRS) and Low Resistance State (LRS).
If a CAM entry value is '1', the RRAM pair is programmed to [HRS,LRS], thus if its input on the Select Line (SL) is also '1' only a small leakage current is generated on the Match Line (ML), given that SL is applied to the leftmost RRAM and SL-negated to the rightmost RRAM.
On the other hand, a '0' is mapped as [LRS,HRS]. 
Multiple 2T2R CAM cells are arranged in a row, inputs are applied along columns, and if the current at the rows is $\sim0$ a match is found, namely the stored word and input are the same.
Note that effectively, the CAM computes the Hamming Distance $\delta(\cdot)$ of the input and the stored word, if $\textit{ML}=\delta(x, T_i)==0$ a match is found.
Finally, if a [HRS, HRS] couple is programmed, the resulting current will be a small leakage both in the case a '1' or '0' is searched, a state known as wildcard 'X', resulting in a Ternary CAM (TCAM).
Fig.~\ref{fig:si:tcam} shows details on mapping, searching, and computing $\delta(\cdot)$ with TCAMs. 

A TCAM can be used to map and evaluate a SAT problem, and computing $\delta(\cdot)$ can give information about the variable interactions in each clause.
Fig. \ref{fig:intro}b shows a circuit implementation of the proposed $k-$Local Interactions In-Memory Accelerator (KLIMA) employing a TCAM for calculating interactions between variables and a second crossbar array in an HNN-like configuration for computing gradients, based on several SLS heuristics.
The design is similar to an HNN, with minimal peripheral circuits consisting of sense amplifiers, noise generators, and logic gates, as shown in the complete schematic Fig.~\ref{fig:si:klima}.

Fig. \ref{fig:klima}c shows an example of using KLIMA to compute the make value $m$ \cite{pedretti2023zeroth}, one of the key SAT-specific metrics used for gradient computation and other operations in state-of-the-art SLS solvers.
The make value corresponds to how many clauses that are currently violated by a given variable, will be satisfied if the variable is flipped.
Note that thanks to the disjunction in each clause, if a clause is violated it is sufficient to flip one of its member variables to satisfy it.
First, the TCAM is used to compute the location of violated clauses.
Each clause is mapped to a TCAM row $T_i$ such as if the Hamming distance between the input $x$ and $T_i$ $D=\delta(x, T_i)=0$, corresponding to a match, the clause is violated.
In the example, the formula of Fig. \ref{fig:intro}a is mapped to the TCAM with positive literals (e.g. $x_1$ in the first clause) encoded as 0, negative literals (such as $x_4$ in the first clause) encoded as 1 and the absence of a literal in the clause mapped as a wildcard X.
By applying the input $x$ on the SLs, the resulting hamming distance is $\text{ML}=\delta(x, T)=[1,0,0]$, with $T$ the TCAM coupling terms.
The current flowing into the TCAM rows $i$ is sensed with a Sense Amplifier (SA) with threshold $\Theta_h=1$ to check where $\text{MLm}=\delta(x, T_i)<\Theta_h$, corresponding to a match, thus the violation of the corresponding clause. 
To compute $m$ it is possible to perform a dot product.
The formula of Fig. \ref{fig:intro}a is mapped to a DPE such as a member variable in a clause is encoded as 1 (such as $x_1$, $x_2$, and $x_4$ in the first clause), while the other variables as 0.
The TCAM MLs after sensing, are directly connected to the DPE WLs, such that activated rows correspond to violated clauses, and respective member variables in such clauses are accumulated on the BLs.
The output on the BLs corresponds to the $m$.

Note that as illustrated by Fig. \ref{fig:klima}d it is possible to compute the break value $b$ with a similar mechanism~\cite{bhattacharya2024computing}. 
The break value of a variable is another key metric of SLS SAT solvers, representing how many clauses currently satisfied will be violated if that variable is flipped.
This time the TCAM is used to compute clauses in which just one literal is satisfied, thus if that literal is flipped the clause becomes unsatisfied.
To do that, it is possible to sense again the Hamming distance $\text{ML}$ to check where $\delta(x, T_i)=1$ through an SA with a double threshold, namely $\Theta_h=2$ and $\Theta_h=0$, such that $\text{MLb}=(\Theta_l<\text{ML}<\Theta_h)$.
Note however, that with $\text{MLb}$ we know that a clause has a single satisfied literal, but we don't know yet which variable is responsible for that.
Thus, to compute $b$, a second crossbar is used with the same mapping as the TCAM.
In this way, the currents generated by a positive or negative literal of a member variable are separated, such that when $\text{MLb}$ is applied to the rows, the individual contribution in each column can be multiplied with the current state $x$, in order to select only variables which are part of the break count.
Given inputs $x$ are binary, a simple transmission gate can be used for propagating the correct break counts.
As shown in Fig. \ref{fig:klima}e, the gain value, namely the difference between make and break, can be computed by $g = m - b$.
Finally, the variable to be flipped, e.g. with the highest gain, is selected with a Winner Takes All (WTA) circuit, after noise is added.
Note that the logical matrix stored in the respective DPEs for computing make and break values, is identical. Furthermore, both $m$ and $b$ (and subsequently $g$ as well) could be computed using a single DPE in a single cycle with slightly modified word line activation protocol \cite{bhattacharya2024computing}.
Furthermore, in principle a single crossbar array with bidirectional operation~\cite{marinella2018multiscale} can perform both the TCAM and DPE operation, resulting in significant area efficiency.

In the KLIMA approach, the Hamming distance is used to evaluate the interactions between clauses without needing to specifically compute higher-order products.
Interestingly, natively mapping the problem in the CNF space without the need for auxiliary variables results in a significant mapping advantage in terms of coupling terms, compared to a Q-HNN.
We define the mapping advantage as
\begin{equation}
    \Sigma = M_{HNN}/M_{KLIMA}
\end{equation}
with $M_{HNN}$ and $M_{KLIMA}$ representing the number of coupling terms for quadratic HNN and KLIMA, respectively.

As an example, we studied $\Sigma$ as a function of the order $k$ for hard $k-$SAT instances at the computational phase transition, comparing KLIMA with a QUBO mapped HNN using the Rosenberg mapping \cite{rosenberg1975reduction,hizzani2024memristor,dobrynin2024energy}.
The computational phase transition is defined as the clause-to-variable ratio $\alpha = C/V$ such that there is a sudden change in the probability of finding a satisfying assignment. 
Below this threshold, most instances are satisfiable, while above it, most are unsatisfiable. 
This behavior resembles physical phase transitions, where properties of a system change dramatically at a specific point~\cite{krzakala2007gibbs}.
Fig. \ref{fig:klima}f shows $\alpha$ at the computational phase transition and $\Sigma$ as a function of $k$.
The coupling terms for KLIMA are computed as
\begin{equation}
    M_{\rm KLIMA} = (2+2)CV = 4 \alpha V^2 = 4 \alpha V^2
\end{equation}
with the factor 4 accounting for the TCAM and DPE with 2T2R cells.
In the case of QUBO-mapped HNN, the number of coupling terms can be computed as
\begin{equation}
    M_{\rm HNN} = 2[V+(k-2)C]^2 = 2[1+(k-2)\alpha]^2V^2
\end{equation}
with the factor 2 accounting for positive and negative weights, and $(k-2)$ for an auxiliary variable added for each clause with $k>2$.
The resulting advantage is
\begin{equation}
    \Sigma = \frac{M_{\rm HNN}}{M_{\rm KLIMA}} =  \frac{2}{4}\frac{[1+(k-2)\alpha]^2}{\alpha} \sim O(k^2\alpha).
\end{equation}
Note that interestingly the number of required resources for KLIMA is independent of $k$ thus the advantage  $\Sigma$ scales aggressively with it, as relevant problems have $k\gg 2$.

\subsection{Heuristic optimization}\label{sec:heuristic}
\begin{figure}
    \centering
    \includegraphics[width=0.7\linewidth]{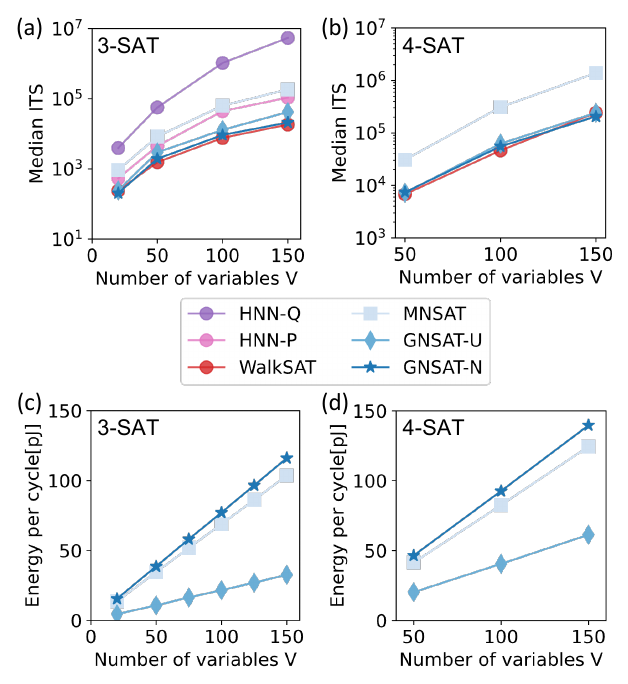}
    \caption{(a) Median ITS as a function of the number of variables for solving 3-SAT and (b) 4-SAT problems with multiple heuristics. (c) Median energy per cycle as a function of the number of variables for multiple KLIMA operations solving 3-SAT and (d) 4-SAT problems.}
    \label{fig:its}
\end{figure}

Given KLIMA's flexibility to compute the make, break, and gain values, we can implement different solvers.
Here, we focus on two basic SLS algorithms namely GSAT and WalkSAT~\cite{hoos2000local}.
GSAT (Algorithm~\ref{si:algo:GSAT}) updates the variable with the highest gain at each step.
However, GSAT may get stuck in local minima.
To perform random walks, in WalkSAT (Algorithm~\ref{si:algo:WalkSAT}) first a random violated clause is selected, then either the clause member variable with the highest gain or a random member variable is flipped.
A hybrid GSAT-WalkSAT heuristic (GWSAT, Algorithm~\ref{si:algo:GWSAT}) was also proposed.
\rev{KLIMA can natively map traditional heuristics such as GWSAT, but it would not take full advantage of the hardware}.
Taking inspiration from the original GSAT, thanks to the possibility of KLIMA to compute the make value of all variables in parallel, we designed a heuristic (MNSAT, Algorithm~\ref{si:algo:MNSAT}) in which at each step, the variable with the highest make value is updated~\cite{pedretti2023zeroth}.
To avoid getting stuck in local minima, but avoiding digital control by a state machine for random walks, we took inspiration from how simulated annealing is implemented in HNNs, by adding random noise to each \rev{incoming current in the neurons}~\cite{cai2020power}.
However, make values contain less information compared to the gain.
Flipping a variable might satisfy (or make) one or more new clauses, but might unsatisfy (or break) one or more other clauses.

\begin{algorithm}
\caption{GNSAT}\label{algo:GNSAT}
\begin{algorithmic}[1]
\renewcommand{\algorithmicrequire}{\textbf{Input:}}
\renewcommand{\algorithmicensure}{\textbf{Output:}}
\Require CNF Formula $y$, Boolean input vector randomly initialized $x_0$, maximum number of iterations $\text{MAX}_{flips}$, noise distribution $\mathcal{N}$, standard deviation of noise $\sigma_N$
\Ensure Input vector $x$ that maximizes the number of satisfied clauses 
\State $t \Leftarrow0$
\State $x(0) \Leftarrow x_0$
\While{$t<\text{MAX}_{flips}$ or $\text{UNSAT}$}
    \State Evaluate the number of violated clauses $w$
    \If{$w==0$}
        \State The problem is satisfied $\text{UNSAT} = False$
    \Else%[$w>0$]
        \State Compute the gain value for all the variables $g$
        \State Add random noise $g' \Leftarrow \mathcal{N}(g,\sigma_N)$
        \State$x(t+1)\Leftarrow x(t)$
        \State $x(t+1)[\text{argmax}(g')] \Leftarrow 1 - x(t+1)[\text{argmax}(g')]$
        \State $t \Leftarrow t+1$        
    \EndIf
\EndWhile
\end{algorithmic} 
\end{algorithm}
 
Algorithm \ref{algo:GNSAT} shows a novel proposed solver,  namely GNSAT in which, similarly to GSAT, the variable with the highest gain among all the violated clauses is updated but random noise is added to the array of gain values before selecting the best one to flip, as in MNSAT~\cite{pedretti2023zeroth}.
We considered two noise profiles $\mathcal{N}$, namely normal distributed noise (GNSAT-N) and uniformly distributed noise (GNSAT-U).

We compared the Iterations To Solution to solve the problem with 99\% probability (ITS) using various SLS solvers, including WalkSAT~\cite{WinNT}, with HNNs solvers based on QUBO and PUBO mapping~\cite{hizzani2024memristor}, namely HNN-Q and HNN-P, on the random uniform 3-SAT \cite{hoos2000satlib} and 4-SAT problems with increasing number of variables and a clause to variable ratio at the computational phase transition.
Fig. \ref{fig:its} shows the Median ITS as a function of the number of variables for 3-SAT (a) and 4-SAT (b).
Note that in the case of 4-SAT we only tested the Walk-based approaches given the prohibitively long simulation times of the HNN solvers.
Details on how the datasets and benchmark experiments were run can be found in Methods.
Results highlight how (1) WalkSAT-based solvers outperform HNNs solvers, (2) the state-of-the-art WalkSAT-SKC algorithm~\cite{WinNT} is comparable with our co-designed solvers, and (3) GNSAT-N outperforms GNSAT-U.

\subsection{Hardware-Software codesign}\label{sec:hwsw}
\begin{figure}
    \centering
    \includegraphics[width=0.7\linewidth]{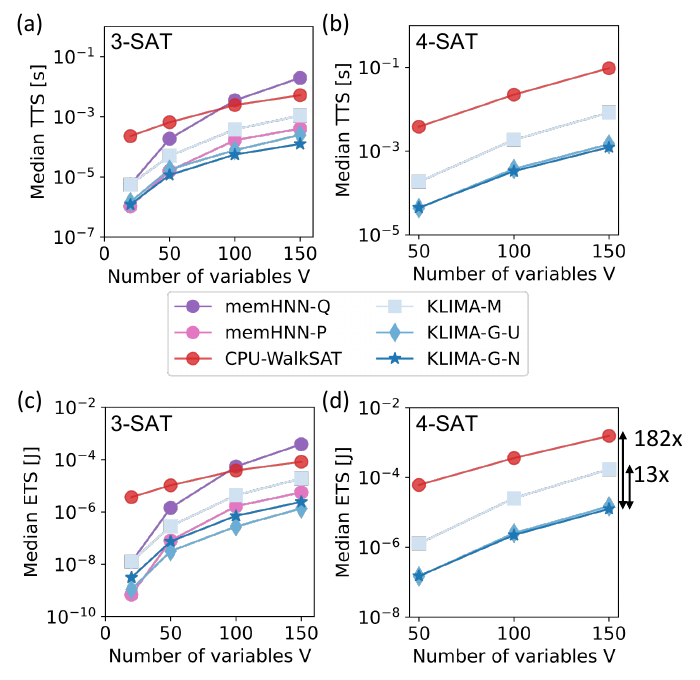}
    \caption{(a) Median TTS as a function of the number of variables for solving 3-SAT and (b) 4-SAT problems with multiple heuristics. (c) Median energy per cycle as a function of the number of variables for multiple KLIMA operations solving 3-SAT and (d) 4-SAT problems.}
    \label{fig:tts-ets}
\end{figure}

The implementation on our KLIMA hardware of the various algorithms results in different energy and time per iteration.
Computing the make value is simple, it only needs a single SA per ML and a single dot product, while the gain value is more complicated, as it also needs the computation of the break value followed by performing the subtraction.
Uniformly generated noise is easy to obtain with digital Pseudo Random Number Generators (PRNGs) based for example on XOR-SHIFT and a Digital Analog Converter (DAC) while implementing a good Gaussian RNG requires significantly more expensive circuitry (Fig.~\ref{fig:si:components}c-d) \rev{which takes most of the energy consumption }(Fig.~\ref{fig:si:breakdown}).
We envision the possibility of using intrinsic RRAM noise~\cite{cai2020power,mahmoodi2019versatile}, although out of the scope of this work.
For these reasons, the ITS alone is not enough to understand which heuristic is more performant.

To assess the impact of various solvers on the overall hardware metrics, namely Time To Solution (TTS) and Energy To Solution (ETS), we carefully designed and characterized the activity of each component involved in the circuit implementation (see Methods).
Fig. \ref{fig:its}c shows the energy per cycle, or iteration, to run different heuristics on KLIMA as a function of the number of variables (see Methods for details) for 3-SAT problems.
Results show a big effect on the energy consumption of the noise generation.
Sampling from a normal distribution is significantly more expensive than performing a VMM on a crossbar array.
Similar results are shown in Fig. \ref{fig:its}d for 4-SAT problems.
Fig. \ref{fig:tts-ets} shows the median TTS for solving 3-SAT (a) and 4-SAT (b) using the following hardware implementations: QUBO and PUBO-HNN implemented using RRAM arrays, namely memHNN-Q \cite{cai2020power} and memHNN-P \cite{hizzani2024memristor,bhattacharya2024computing} respectively; WalkSAT on CPU (CPU-WalkSAT), MNSAT ~\cite{pedretti2023zeroth}, GNSAT-U and GNSAT-N implemented using our KLIMA hardware, referred to as KLIMA-M, KLIMA-G-U and KLIMA-G-N respectively. Similarly, Fig. \ref{fig:tts-ets}c,d shows the median ETS for solving 3-SAT and 4-SAT respectively.
Results demonstrate the higher performance of KLIMA accelerated solvers over both memHNN implementations and CPU running WalkSAT.
Moreover, both variants of KLIMA-G (-U and N respectively) outperform KLIMA-M significantly, while the noise profile seems to have only minimal impact.

\section{Discussion}\label{sec:discussion}

We presented KLIMA, an in-memory computing SAT solver accelerator that uses TCAMs for encoding higher-order interactions in the native problem space without the need for auxiliary variables.
Native mapping~\cite{pedretti2023zeroth,bhattacharya2024computing} allows for a better representation of the problem, without artificial local minima and saddle points~\cite{dobrynin2024energy} and significantly reduced hardware resources.
Interestingly, KLIMA enables highly parallel implementation of a wide range of SLS-based SAT-solving heuristics that are commonly used in literature as well as novel variants. This versatility allows hardware-software codesign to develop solvers that are optimized for the given type of SAT instances and meet the latency, power, and area budget of the application.
We designed multiple KLIMA-compatible heuristics and the corresponding peripherals to pick the most energy-efficient for random uniform SAT instances, in which SLS solvers shine.
Results show 182$\times$ improvement compared with the state-of-the-art digital implementation on CPU and 14$\times$ compared to our previously presented results \cite{pedretti2023zeroth} for random uniform 4-SAT problem, suggesting even higher performance with higher $k$.

The proposed in-memory massively parallel $k-$local interaction computation and the presented heuristic exploration is an initial step towards a novel class of computing primitives, solver heuristics, and potentially machine learning models~\cite{niazi2024training}.
We envision KLIMA as a building block for application-specific and/or heterogenous~\cite{mohseni2024build} computing systems for optimization and sampling, removing previous bottlenecks and leading to unprecedented performance.

\section{Methods}\label{sec:methods}
\subsection{Heuristic simulations}
Heuristics were simulated using CountryCrab, a distributed simulator for optimization solvers based on analog hardware. 
CountryCrab is based on Python and built on CuPy~\cite{nishino2017cupy}.
Each instance initializes a $V \times\text{MAX}_{tries}$ matrix representing the candidate solutions and a $V\times C\times \text{MAX}_{tries}$ matrix representing the problem, which is processed for $\text{MAX}_{flips}$ iterations to simulate the hardware operation. 
CountryCrab allows matrices representing the problem to have any dimension, including for example the number of cores, number of tiles, number of chips, and so on, resulting in a highly scalable tool.
Instances are parallelized using Multiprocessing with each process allocating a certain amount of GPU memory.
MLflow~\cite{zaharia2018accelerating} and Ray Tune~\cite{liaw2018tune} are used for scheduling, tracking, and logging experiments.
We considered uniformly distributed random problems from the SATLIB dataset~\cite{hoos2000satlib} in the case of 3-SAT and generated similarly difficult problems in the case of 4-SAT.

\subsection{Hyperparameter optimization}
To optimize hyperparameters such as the amount of noise and the $\text{MAX}_{flips}$ before restarting the solvers, we used 20\% of the instances for each size and tried to solve them for multiple hyperparameters.
In the case of KLIMA, we randomly sampled 20 relative standard deviations of noise from a uniform distribution and solved 20\% of instances until convergence or for a large number of iterations, such as 50E3.
For each instance we repeated the experiments $\text{MAX}_{tries}=1000$ tries and then we computed the ITS at each iteration $t$ 
\begin{equation}
    ITS(t) = t\cdot \frac{log(1-P_{target})}{log(1-P(t))} 
\end{equation}
with $P_{target}=0.99$ the target probability to solve the problem, and $P(t)$ the probability to solve it at iteration $t$, namely the number of solved tries normalized by $\text{MAX}_{tries}$
For each instance, we picked the optimized $\text{MAX}_{flips}=\text{argmin}_t(ITS(t))$ and found the corresponding best noise.
We then computed the median value of $\text{MAX}_{flips}$ over the instances and the median relative standard deviation of noise to be used for benchmarking in the remaining 80\% of instances.

\subsection{KLIMA energy modeling}
We simulated the energy consumption and latency of the crossbar array and peripheral circuit of KLIMA in 28 nm TSMC (details in Methods and Supplementary Information~\ref{si:sec-modeling} and Fig.~\ref{fig:si:components}).
The latency was modeled as $t_{clk} = 6 \,$ns and we calculated the time to solution as $TTS = t_{clk}\cdot ITS$.
RRAM data is modeled from previously fabricated test chips~\cite{sheng2019low}.
We build a custom KLIMA event-based simulator which allows us to get detailed information on the utilization of each component, without extensive spice-like simulations which would be intractable.
We simulated the mean energy consumption $E_{mean}$ for each instance and calculated the energy to solution as $ETS = E_{mean}\cdot ITS$

\subsection{CPU energy modeling}
Since KLIMA results are based on simulations, we simulated a CPU using a similar approach.
We used the Ariel core in SST to simulate the resource utilization, specifically the memory access which accounts for most of the energy consumption.
We then used CACTI~\cite{muralimanohar2009cacti} for estimating the memory energy consumption from its utilization.
We based the CPU architecture on the Intel Xeon CPU E5-4669 v3 on an HPE ProLiant DL560Gen9 server with a bus latency of $50\,\mathrm{ps}$, cache frequency of 2.1$\,$GHz, and cache sizes of 32$\,$KB (with associativity = 8), 256$\,$KB (with associativity = 8), and 188$\,$MB (with associativity = 30) for L1, L2, and L3 respectively.
We experimentally collect ITS data from a state-of-the-art WalkSAT-SKC solver~\cite{WinNT}, optimizing the hyperparameters in the same way as KLIMA, and compute TTS and ETS with data from the CPU simulations.

Note that this is a lower bound for the CPU energy consumption, which does not include any communication between memory and CPU, nor computations in the CPU.
Thus, we believe that the simulation is fair, and in practice, KLIMA may have an even larger advantage over the CPU-based solvers.

\backmatter

\bmhead{Supplementary information}
Supplementary information can be found in the Supplementary Information Document

\bmhead{Acknowledgements}
This work is supported by the Defense Advanced Research Projects Agency (DARPA) under Air Force Research Laboratory (AFRL) contract no FA8650-23-3-7313.

\section*{Declarations}
\begin{itemize}
\item \textbf{Competing interests}:\\
The authors declare no competing interests.
% \item \textbf{Data availability}:\\
% The data that support the findings in this paper are provided in the main text, Supplementary Information file and available code repository. Additional data related to this study can be made available from the corresponding authors upon request.
\item \textbf{Code availability}:\\
The simulator used for heuristic estimation is open-sourced and available at \url{https://github.com/HewlettPackard/CountryCrab}
% \item \textbf{Author contribution}\\
% TBD
\end{itemize}

\bibliography{sn-bibliography}% common bib file

\begin{thebibliography}{10}
\expandafter\ifx\csname url\endcsname\relax
  \def\url#1{\burl{#1}}\fi
\expandafter\ifx\csname urlprefix\endcsname\relax\def\urlprefix{URL }\fi
\providecommand{\bibinfo}[2]{#2}
\providecommand{\eprint}[2][]{\url{#2}}
\providecommand{\doi}[1]{\url{https://doi.org/#1}}
\bibcommenthead

\bibitem{marques2008practical}
\bibinfo{author}{Marques-Silva, J.}
\newblock \bibinfo{title}{Practical applications of boolean satisfiability}.
\newblock \emph{\bibinfo{journal}{2008 9th International Workshop on Discrete Event Systems}} \bibinfo{pages}{74--80} (\bibinfo{year}{2008}).

\bibitem{burch1992symbolic}
\bibinfo{author}{Burch, J.~R.}, \bibinfo{author}{Clarke, E.~M.}, \bibinfo{author}{McMillan, K.~L.}, \bibinfo{author}{Dill, D.~L.} \& \bibinfo{author}{Hwang, L.-J.}
\newblock \bibinfo{title}{Symbolic model checking: 1020 states and beyond}.
\newblock \emph{\bibinfo{journal}{Information and computation}} \textbf{\bibinfo{volume}{98}}, \bibinfo{pages}{142--170} (\bibinfo{year}{1992}).

\bibitem{rintanen2011planning}
\bibinfo{author}{Rintanen, J.}
\newblock \bibinfo{title}{Planning with specialized sat solvers}.
\newblock \emph{\bibinfo{journal}{Proceedings of the AAAI Conference on Artificial Intelligence}} \textbf{\bibinfo{volume}{25}}, \bibinfo{pages}{1563--1566} (\bibinfo{year}{2011}).

\bibitem{horbach2012using}
\bibinfo{author}{Horbach, A.}, \bibinfo{author}{Bartsch, T.} \& \bibinfo{author}{Briskorn, D.}
\newblock \bibinfo{title}{Using a sat-solver to schedule sports leagues}.
\newblock \emph{\bibinfo{journal}{Journal of Scheduling}} \textbf{\bibinfo{volume}{15}}, \bibinfo{pages}{117--125} (\bibinfo{year}{2012}).

\bibitem{trinh2024solving}
\bibinfo{author}{Trinh, T.~H.}, \bibinfo{author}{Wu, Y.}, \bibinfo{author}{Le, Q.~V.}, \bibinfo{author}{He, H.} \& \bibinfo{author}{Luong, T.}
\newblock \bibinfo{title}{Solving olympiad geometry without human demonstrations}.
\newblock \emph{\bibinfo{journal}{Nature}} \textbf{\bibinfo{volume}{625}}, \bibinfo{pages}{476--482} (\bibinfo{year}{2024}).

\bibitem{yang2024fine}
\bibinfo{author}{Yang, Y.} \emph{et~al.}
\newblock \bibinfo{title}{Fine-tuning language models using formal methods feedback: A use case in autonomous systems}.
\newblock \emph{\bibinfo{journal}{Proceedings of Machine Learning and Systems}} \textbf{\bibinfo{volume}{6}}, \bibinfo{pages}{339--350} (\bibinfo{year}{2024}).

\bibitem{hoos2000local}
\bibinfo{author}{Hoos, H.~H.} \& \bibinfo{author}{St{\"u}tzle, T.}
\newblock \bibinfo{title}{Local search algorithms for sat: An empirical evaluation}.
\newblock \emph{\bibinfo{journal}{Journal of Automated Reasoning}} \textbf{\bibinfo{volume}{24}}, \bibinfo{pages}{421--481} (\bibinfo{year}{2000}).

\bibitem{davis1962machine}
\bibinfo{author}{Davis, M.}, \bibinfo{author}{Logemann, G.} \& \bibinfo{author}{Loveland, D.}
\newblock \bibinfo{title}{A machine program for theorem-proving}.
\newblock \emph{\bibinfo{journal}{Communications of the ACM}} \textbf{\bibinfo{volume}{5}}, \bibinfo{pages}{394--397} (\bibinfo{year}{1962}).

\bibitem{cai2021deep}
\bibinfo{author}{Cai, S.} \& \bibinfo{author}{Zhang, X.}
\newblock \bibinfo{title}{Deep cooperation of cdcl and local search for sat}.
\newblock \emph{\bibinfo{journal}{Theory and Applications of Satisfiability Testing--SAT 2021: 24th International Conference, Barcelona, Spain, July 5-9, 2021, Proceedings 24}} \bibinfo{pages}{64--81} (\bibinfo{year}{2021}).

\bibitem{mohseni2022ising}
\bibinfo{author}{Mohseni, N.}, \bibinfo{author}{McMahon, P.~L.} \& \bibinfo{author}{Byrnes, T.}
\newblock \bibinfo{title}{Ising machines as hardware solvers of combinatorial optimization problems}.
\newblock \emph{\bibinfo{journal}{Nature Reviews Physics}} \textbf{\bibinfo{volume}{4}}, \bibinfo{pages}{363--379} (\bibinfo{year}{2022}).

\bibitem{hopfield1985neural}
\bibinfo{author}{Hopfield, J.~J.} \& \bibinfo{author}{Tank, D.~W.}
\newblock \bibinfo{title}{“neural” computation of decisions in optimization problems}.
\newblock \emph{\bibinfo{journal}{Biological cybernetics}} \textbf{\bibinfo{volume}{52}}, \bibinfo{pages}{141--152} (\bibinfo{year}{1985}).

\bibitem{cai2020power}
\bibinfo{author}{Cai, F.} \emph{et~al.}
\newblock \bibinfo{title}{Power-efficient combinatorial optimization using intrinsic noise in memristor hopfield neural networks}.
\newblock \emph{\bibinfo{journal}{Nature Electronics}} \textbf{\bibinfo{volume}{3}}, \bibinfo{pages}{409--418} (\bibinfo{year}{2020}).

\bibitem{mahmoodi2019versatile}
\bibinfo{author}{Mahmoodi, M.}, \bibinfo{author}{Prezioso, M.} \& \bibinfo{author}{Strukov, D.}
\newblock \bibinfo{title}{Versatile stochastic dot product circuits based on nonvolatile memories for high performance neurocomputing and neurooptimization}.
\newblock \emph{\bibinfo{journal}{Nature communications}} \textbf{\bibinfo{volume}{10}}, \bibinfo{pages}{5113} (\bibinfo{year}{2019}).

\bibitem{aadit2022massively}
\bibinfo{author}{Aadit, N.~A.} \emph{et~al.}
\newblock \bibinfo{title}{Massively parallel probabilistic computing with sparse ising machines}.
\newblock \emph{\bibinfo{journal}{Nature Electronics}} \textbf{\bibinfo{volume}{5}}, \bibinfo{pages}{460--468} (\bibinfo{year}{2022}).

\bibitem{bohm2019poor}
\bibinfo{author}{B{\"o}hm, F.}, \bibinfo{author}{Verschaffelt, G.} \& \bibinfo{author}{Van~der Sande, G.}
\newblock \bibinfo{title}{A poor man’s coherent ising machine based on opto-electronic feedback systems for solving optimization problems}.
\newblock \emph{\bibinfo{journal}{Nature communications}} \textbf{\bibinfo{volume}{10}}, \bibinfo{pages}{3538} (\bibinfo{year}{2019}).

\bibitem{ielmini2018memory}
\bibinfo{author}{Ielmini, D.} \& \bibinfo{author}{Wong, H.-S.~P.}
\newblock \bibinfo{title}{In-memory computing with resistive switching devices}.
\newblock \emph{\bibinfo{journal}{Nature electronics}} \textbf{\bibinfo{volume}{1}}, \bibinfo{pages}{333--343} (\bibinfo{year}{2018}).

\bibitem{ielmini2016resistive}
\bibinfo{author}{Ielmini, D.}
\newblock \bibinfo{title}{Resistive switching memories based on metal oxides: mechanisms, reliability and scaling}.
\newblock \emph{\bibinfo{journal}{Semiconductor Science and Technology}} \textbf{\bibinfo{volume}{31}}, \bibinfo{pages}{063002} (\bibinfo{year}{2016}).

\bibitem{burr2016recent}
\bibinfo{author}{Burr, G.~W.} \emph{et~al.}
\newblock \bibinfo{title}{Recent progress in phase-change memory technology}.
\newblock \emph{\bibinfo{journal}{IEEE Journal on Emerging and Selected Topics in Circuits and Systems}} \textbf{\bibinfo{volume}{6}}, \bibinfo{pages}{146--162} (\bibinfo{year}{2016}).

\bibitem{ielmini2020device}
\bibinfo{author}{Ielmini, D.} \& \bibinfo{author}{Pedretti, G.}
\newblock \bibinfo{title}{Device and circuit architectures for in-memory computing}.
\newblock \emph{\bibinfo{journal}{Advanced Intelligent Systems}} \textbf{\bibinfo{volume}{2}}, \bibinfo{pages}{2000040} (\bibinfo{year}{2020}).

\bibitem{hizzani2024memristor}
\bibinfo{author}{Hizzani, M.} \emph{et~al.}
\newblock \bibinfo{title}{Memristor-based hardware and algorithms for higher-order hopfield optimization solver outperforming quadratic ising machines}.
\newblock \emph{\bibinfo{journal}{2024 IEEE International Symposium on Circuits and Systems (ISCAS)}} \bibinfo{pages}{1--5} (\bibinfo{year}{2024}).

\bibitem{bhattacharya2024computing}
\bibinfo{author}{Bhattacharya, T.} \emph{et~al.}
\newblock \bibinfo{title}{Computing high-degree polynomial gradients in memory}.
\newblock \emph{\bibinfo{journal}{Nature Communications}} \textbf{\bibinfo{volume}{15}}, \bibinfo{pages}{8211} (\bibinfo{year}{2024}).

\bibitem{pedretti2022differentiable}
\bibinfo{author}{Pedretti, G.} \emph{et~al.}
\newblock \bibinfo{title}{Differentiable content addressable memory with memristors}.
\newblock \emph{\bibinfo{journal}{Advanced electronic materials}} \textbf{\bibinfo{volume}{8}}, \bibinfo{pages}{2101198} (\bibinfo{year}{2022}).

\bibitem{pedretti2023zeroth}
\bibinfo{author}{Pedretti, G.} \emph{et~al.}
\newblock \bibinfo{title}{Zeroth and higher-order logic with content addressable memories}.
\newblock \emph{\bibinfo{journal}{2023 International Electron Devices Meeting (IEDM)}} \bibinfo{pages}{1--4} (\bibinfo{year}{2023}).

\bibitem{chiang2024reaim}
\bibinfo{author}{Chiang, H.-W.}, \bibinfo{author}{Nien, C.-F.}, \bibinfo{author}{Cheng, H.-Y.} \& \bibinfo{author}{Huang, K.-P.}
\newblock \bibinfo{title}{Reaim: A reram-based adaptive ising machine for solving combinatorial optimization problems}.
\newblock \emph{\bibinfo{journal}{2024 ACM/IEEE 51st Annual International Symposium on Computer Architecture (ISCA)}} \bibinfo{pages}{58--72} (\bibinfo{year}{2024}).

\bibitem{kirkpatrick1983optimization}
\bibinfo{author}{Kirkpatrick, S.}, \bibinfo{author}{Gelatt~Jr, C.~D.} \& \bibinfo{author}{Vecchi, M.~P.}
\newblock \bibinfo{title}{Optimization by simulated annealing}.
\newblock \emph{\bibinfo{journal}{Science}} \textbf{\bibinfo{volume}{220}}, \bibinfo{pages}{671--680} (\bibinfo{year}{1983}).

\bibitem{lucas2014ising}
\bibinfo{author}{Lucas, A.}
\newblock \bibinfo{title}{Ising formulations of many np problems}.
\newblock \emph{\bibinfo{journal}{Frontiers in physics}} \textbf{\bibinfo{volume}{2}}, \bibinfo{pages}{5} (\bibinfo{year}{2014}).

\bibitem{valiante2021computational}
\bibinfo{author}{Valiante, E.}, \bibinfo{author}{Hernandez, M.}, \bibinfo{author}{Barzegar, A.} \& \bibinfo{author}{Katzgraber, H.~G.}
\newblock \bibinfo{title}{Computational overhead of locality reduction in binary optimization problems}.
\newblock \emph{\bibinfo{journal}{Computer Physics Communications}} \textbf{\bibinfo{volume}{269}}, \bibinfo{pages}{108102} (\bibinfo{year}{2021}).

\bibitem{dobrynin2024energy}
\bibinfo{author}{Dobrynin, D.} \emph{et~al.}
\newblock \bibinfo{title}{Energy landscapes of combinatorial optimization in ising machines}.
\newblock \emph{\bibinfo{journal}{Phys. Rev. E}}  (\bibinfo{year}{2024}).

\bibitem{sharma2023augmenting}
\bibinfo{author}{Sharma, A.}, \bibinfo{author}{Burns, M.}, \bibinfo{author}{Hahn, A.} \& \bibinfo{author}{Huang, M.}
\newblock \bibinfo{title}{Augmenting an electronic ising machine to effectively solve boolean satisfiability}.
\newblock \emph{\bibinfo{journal}{Scientific Reports}} \textbf{\bibinfo{volume}{13}}, \bibinfo{pages}{22858} (\bibinfo{year}{2023}).

\bibitem{bybee2023efficient}
\bibinfo{author}{Bybee, C.} \emph{et~al.}
\newblock \bibinfo{title}{Efficient optimization with higher-order ising machines}.
\newblock \emph{\bibinfo{journal}{Nature Communications}} \textbf{\bibinfo{volume}{14}}, \bibinfo{pages}{6033} (\bibinfo{year}{2023}).

\bibitem{bhattacharya2024unified}
\bibinfo{author}{Bhattacharya, T.}, \bibinfo{author}{Hutchinson, G.~H.} \& \bibinfo{author}{Strukov, D.~B.}
\newblock \bibinfo{title}{Unified framework for efficient high-order ising machine hardware implementations}.
\newblock \emph{\bibinfo{journal}{Under Review}}  (\bibinfo{year}{2024}).

\bibitem{graves2022memory}
\bibinfo{author}{Graves, C.~E.}, \bibinfo{author}{Li, C.}, \bibinfo{author}{Pedretti, G.} \& \bibinfo{author}{Strachan, J.~P.}
\newblock \bibinfo{title}{In-memory computing with non-volatile memristor cam circuits}.
\newblock \emph{\bibinfo{journal}{Memristor Computing Systems}} \bibinfo{pages}{105--139} (\bibinfo{year}{2022}).

\bibitem{chen2015efficient}
\bibinfo{author}{Chen, B.} \emph{et~al.}
\newblock \bibinfo{title}{Efficient in-memory computing architecture based on crossbar arrays}.
\newblock \emph{\bibinfo{journal}{2015 IEEE International Electron Devices Meeting (IEDM)}} \bibinfo{pages}{17--5} (\bibinfo{year}{2015}).

\bibitem{marinella2018multiscale}
\bibinfo{author}{Marinella, M.~J.} \emph{et~al.}
\newblock \bibinfo{title}{Multiscale co-design analysis of energy, latency, area, and accuracy of a reram analog neural training accelerator}.
\newblock \emph{\bibinfo{journal}{IEEE Journal on Emerging and Selected Topics in Circuits and Systems}} \textbf{\bibinfo{volume}{8}}, \bibinfo{pages}{86--101} (\bibinfo{year}{2018}).

\bibitem{rosenberg1975reduction}
\bibinfo{author}{Rosenberg, I.~G.}
\newblock \bibinfo{title}{Reduction of bivalent maximization to the quadratic case.}
\newblock \emph{\bibinfo{journal}{Cahiers du Centre d’Etudes de Recherche Operationnelle}} \textbf{\bibinfo{volume}{17}}, \bibinfo{pages}{71--79} (\bibinfo{year}{1975}).

\bibitem{krzakala2007gibbs}
\bibinfo{author}{Krzaka{\l}a, F.}, \bibinfo{author}{Montanari, A.}, \bibinfo{author}{Ricci-Tersenghi, F.}, \bibinfo{author}{Semerjian, G.} \& \bibinfo{author}{Zdeborov{\'a}, L.}
\newblock \bibinfo{title}{Gibbs states and the set of solutions of random constraint satisfaction problems}.
\newblock \emph{\bibinfo{journal}{Proceedings of the National Academy of Sciences}} \textbf{\bibinfo{volume}{104}}, \bibinfo{pages}{10318--10323} (\bibinfo{year}{2007}).

\bibitem{WinNT}
\bibinfo{title}{Walksat}.
\newblock \bibinfo{howpublished}{\url{https://gitlab.com/HenryKautz/Walksat}}.
\newblock \bibinfo{note}{Accessed: 2024-10-03}.

\bibitem{hoos2000satlib}
\bibinfo{author}{Hoos, H.~H.} \& \bibinfo{author}{St{\"u}tzle, T.}
\newblock \bibinfo{title}{{SATLIB}: An online resource for research on sat}.
\newblock \emph{\bibinfo{journal}{Sat}} \textbf{\bibinfo{volume}{2000}}, \bibinfo{pages}{283--292} (\bibinfo{year}{2000}).

\bibitem{niazi2024training}
\bibinfo{author}{Niazi, S.} \emph{et~al.}
\newblock \bibinfo{title}{Training deep boltzmann networks with sparse ising machines}.
\newblock \emph{\bibinfo{journal}{Nature Electronics}} \bibinfo{pages}{1--10} (\bibinfo{year}{2024}).

\bibitem{mohseni2024build}
\bibinfo{author}{Mohseni, M.} \emph{et~al.}
\newblock \bibinfo{title}{How to build a quantum supercomputer: Scaling challenges and opportunities}.
\newblock \emph{\bibinfo{journal}{arXiv preprint arXiv:2411.10406}}  (\bibinfo{year}{2024}).

\bibitem{nishino2017cupy}
\bibinfo{author}{Nishino, R.} \& \bibinfo{author}{Loomis, S. H.~C.}
\newblock \bibinfo{title}{Cupy: A numpy-compatible library for nvidia gpu calculations}.
\newblock \emph{\bibinfo{journal}{31st confernce on neural information processing systems}} \textbf{\bibinfo{volume}{151}} (\bibinfo{year}{2017}).

\bibitem{zaharia2018accelerating}
\bibinfo{author}{Zaharia, M.} \emph{et~al.}
\newblock \bibinfo{title}{Accelerating the machine learning lifecycle with mlflow.}
\newblock \emph{\bibinfo{journal}{IEEE Data Eng. Bull.}} \textbf{\bibinfo{volume}{41}}, \bibinfo{pages}{39--45} (\bibinfo{year}{2018}).

\bibitem{liaw2018tune}
\bibinfo{author}{Liaw, R.} \emph{et~al.}
\newblock \bibinfo{title}{Tune: A research platform for distributed model selection and training}.
\newblock \emph{\bibinfo{journal}{arXiv preprint arXiv:1807.05118}}  (\bibinfo{year}{2018}).

\bibitem{sheng2019low}
\bibinfo{author}{Sheng, X.} \emph{et~al.}
\newblock \bibinfo{title}{Low-conductance and multilevel cmos-integrated nanoscale oxide memristors}.
\newblock \emph{\bibinfo{journal}{Advanced electronic materials}} \textbf{\bibinfo{volume}{5}}, \bibinfo{pages}{1800876} (\bibinfo{year}{2019}).

\bibitem{muralimanohar2009cacti}
\bibinfo{author}{Muralimanohar, N.}, \bibinfo{author}{Balasubramonian, R.} \& \bibinfo{author}{Jouppi, N.~P.}
\newblock \bibinfo{title}{Cacti 6.0: A tool to model large caches}.
\newblock \emph{\bibinfo{journal}{HP laboratories}} \textbf{\bibinfo{volume}{27}}, \bibinfo{pages}{28} (\bibinfo{year}{2009}).

\bibitem{walker1974new}
\bibinfo{author}{Walker, A.~J.}
\newblock \bibinfo{title}{New fast method for generating discrete random numbers with arbitrary frequency distributions}.
\newblock \emph{\bibinfo{journal}{Electronics Letters}} \textbf{\bibinfo{volume}{8}}, \bibinfo{pages}{127--128} (\bibinfo{year}{1974}).

\end{thebibliography}
\clearpage

\section{Supplementary information}
\renewcommand{\figurename}{Supplementary Fig.}
\setcounter{figure}{0}
\renewcommand{\thesection}{S.\Roman{section}} 
\renewcommand{\thesubsection}
{\thesection.\Alph{subsection}}
\setcounter{section}{0}
\renewcommand{\thefigure}{S\arabic{figure}}
\renewcommand{\thetable}{S\arabic{table}}
\renewcommand{\thealgorithm}{S\arabic{algorithm}}
\setcounter{algorithm}{0}

\begin{figure}[h!]
    \centering
    \includegraphics[width=0.99\linewidth]{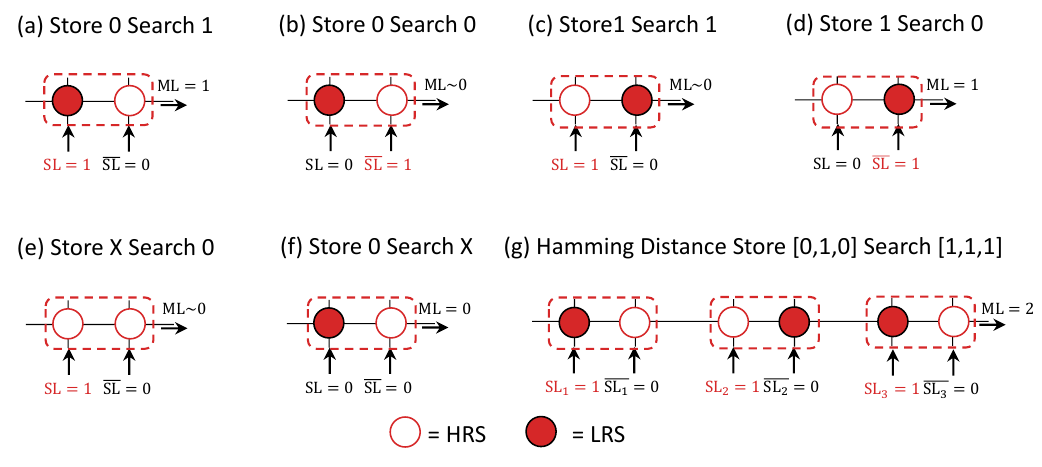}
    \caption{Conceptual operation of TCAM implemented with a crossbar array for multiple search (a,b,c,d,e,f) and compute (g) operations.}
    \label{fig:si:tcam}
\end{figure}

\begin{figure}[h!]
    \centering
    \includegraphics[width=0.65\linewidth]{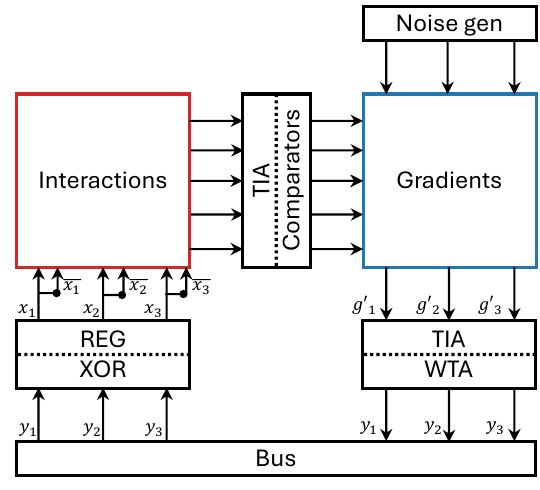}
    \caption{Detailed schematic of KLIMA including all the modeled circuit blocks}
    \label{fig:si:klima}
\end{figure}

\begin{figure}[h!]
    \centering
    \includegraphics[width=0.85\linewidth]{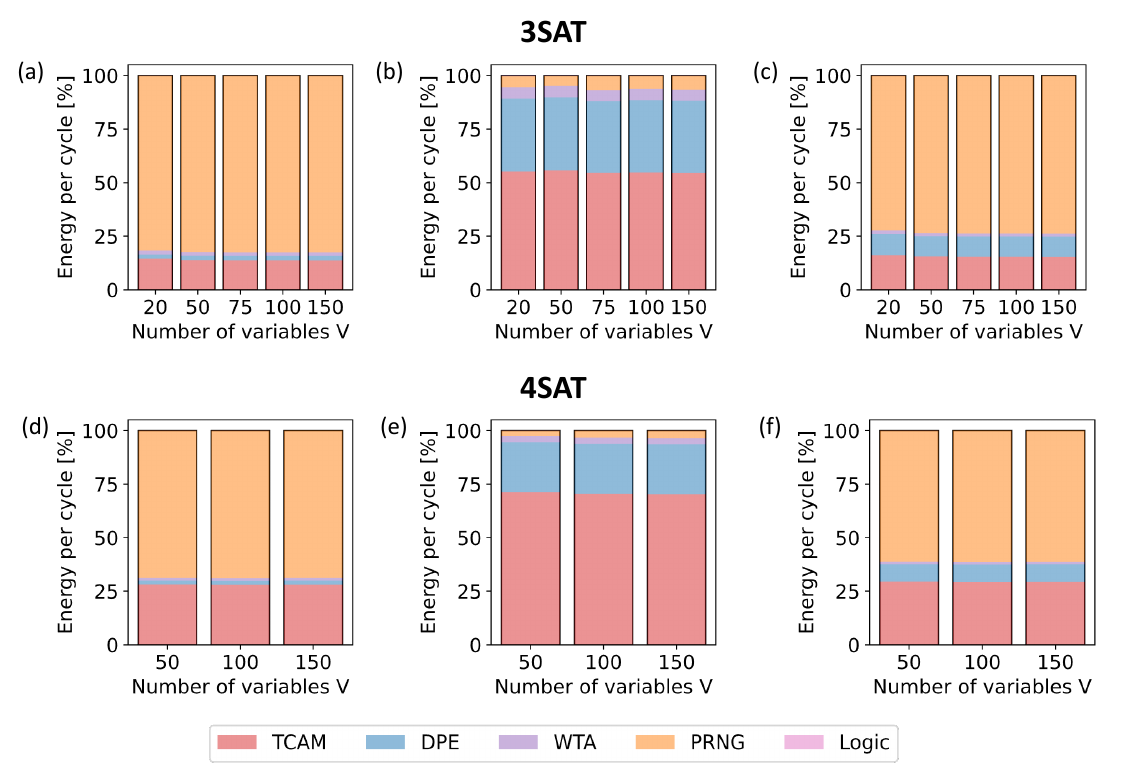}
    \caption{Breakdown of energy consumption for (a,d) KLIMA-M, (b,e) KLIMA-G-U and (c,f) KLIMA-G-N solving 3-SAT and 4-SAT respectively}
    \label{fig:si:breakdown}
\end{figure}

\clearpage

\section{Energy modeling}\label{si:sec-modeling}
\begin{figure}[h!]
    \centering
    \includegraphics[width=0.99\linewidth]{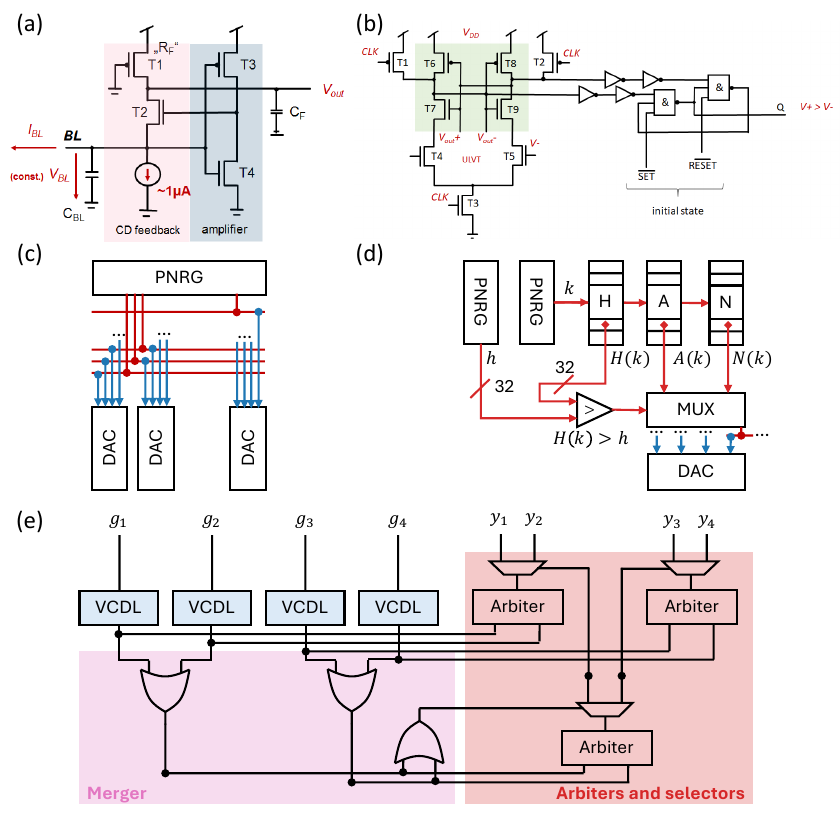}
    \caption{(a) Circuit schematic of Transimpedence Amplifier (TIA). (b) Circuit schematic of the comparator. (c) Block diagram of the uniform random number generator and (d) Normal distributed random number generator. (e) Block diagram of the Winner Takes All (WTA) circuit for computing $\text{argmax}$ function}
    \label{fig:si:components}
\end{figure}

\begin{table}
    \centering
    \begin{tabular}{|c|c|c|c|}
    \hline
         Parameter      & Description                           & Value         & Unit\\
    \hline
         $V_{DD}$       & Supply voltage                        & 0.9           & V\\
         $t_{clk}$      & Clock cycle                           & 2             & ns\\
         $V_{read}$     & Read voltage                          & 0.3           & V\\
         $R_{LRS}$      & RRAM LRS                              & 500E3         & $\Omega$ \\
         $I_{leak}$     & Leakage current in crossbar array     & 2.07          & nA\\
         $I_{TIA,bias}$ & TIA bias current                      & 2             & $\mu$A\\
         $C_{inv}$      & WL inverter input capacitance         & 0.35          & fF\\
         $C_w$          & Wire parasitic capacitance            & 0.22          & fF/$\mu$m\\
         $C_{G}$        & 1T1R gate capacitance                 & 0.3           & fF\\
         $W_{cell}$     & Crossbar cell dimension               & 405           & nm\\
         $E_{XOR}$      & XOR energy consumption                & 1.84          & fJ\\
         $P_{leak, XOR}$& XOR leakage power                     & 7.2           & nW\\
         $E_{REG}$      & Registers energy consumption          & 7.16          & fJ\\
         $P_{leak, REG}$& Registers leakage power               & 41            & nW\\
         $E_{comp}$     & Comparator energy consumption         & 5.5           & fJ\\
         $E_{CLK}$      & Clock tree energy consumption         & 1.85          & fJ\\
         $E_{PNRG}$     & PNRG energy consumption               & 365           & fJ\\
         $E_{VCDL}$     & VCDL energy consumption               & 7.9           & fJ\\
         $E_{WTA,logic}$& WTA logic energy consumption          & 3.4           & fJ\\
         $E_{GPRNG,LUT}$ & GPRNG LUTs energy consumption         & 11.12         & fJ\\
         $E_{GPRNG,comp}$& GPRNG integer comparator energy consumption     & 28         & fJ\\         
    \hline
    \end{tabular}
    \caption{Paramaters for the energy model}
    \label{si:tab:params}
\end{table}

As an example, we explain in detail how the energy consumption for GNSAT is computed, and the same assumptions can be extended to other heuristics.
Fig.~\ref{fig:si:klima} shows the detailed circuit schematic of KLIMA, that we modeled for our co-design effort.
All the circuits were designed, validated, and modeled in TSMC 28 nm~\cite{hizzani2024memristor}.
All the parameters can be found in Table~\ref{si:tab:params}.
We considered 28 nm Back End of the Line (BEOL) integrated ReRAM device based on TaOx~\cite{sheng2019low}. 
To calculate the energy consumption of the crossbar array we first evaluate the mean output levels $\alpha_{ML}$ and $\alpha_{BL}$ with the heuristic simulation (see Methods). 
We evaluate the power consumption of each crossbar column as $P_{col}=\alpha_{x}\frac{V_{DD}V_{read}}{R_{LRS}}$, with $\alpha_{x}$ either $\alpha_{ML}$ or $\alpha_{BL}$ depending if the interaction or gradient crossbar is considered.
Note that the effect of the leakage ($R_{HRS}$) was neglected.
The power consumption to drive each crossbar row is modeled as $P_{row} = V_{DD}I_{leak}(1+\sqrt{C_{row}/C_{inv}})$, with the row capacitance $C_{row}=2N_{cols}(W_{cell}C_w + C_{G})$ dependent on the number of columns $N_{cols}$.
Fig.~\ref{fig:si:components}a shows the circuit schematic of the designed Transimpedence Amplifier (TIA).
Current comes from the DPE columns and is accumulated on a capacitor which drives as pulldown FET (T4).
The current flowing in T4 turns on and consequently T2, thus raising the output voltage.
The power consumption of the TIA is $P_{TIA}=V_{DD}I_{TIA,bias}$.
Thus the energy consumption for the crossbar array is computed as
\begin{equation}
    E_{crossbar}= \alpha_{x}C_{row}^{SW}V_{DD}^2+3t_{CLK}(N_{cols}(P_{row}+P_{TIA})+N_{rows}P_{row})
\end{equation}
with $C_{row}^{SW} = C_{row}+2\sqrt{C_{row}C_{inv}}+2C_{inv}$ the average switching capacitance and the factor $3t_{CLK}$ takes into account the 3 cycles required to operate KLIMA.
Fig.~\ref{fig:si:components}b shows the circuit schematic of the comparator, which is based on a conventional differential structure followed by a pull-up logic.
Comparators are used at the output of the TIA to check, for example, if there is a match or if a clause has only one satisfied literal.
Comparators' energy consumption is shown in Table~\ref{si:tab:params}.
Note that ADCs are not needed during the KLIMA solver operation, and while they could be required for programming purposes, they were not considered in the overall energy consumption.

Fig.~\ref{fig:si:components}c shows the circuit schematic of the uniform noise generation circuit.
A 64-bit pseudorandom number generator (PNRG) based on an XORSHIFT circuit, is used to generate the input to $n_{bDAC}=4$ bit Digital to Analog Converters (DACs).
To save resources, the pseudorandom bits are shared among multiple DACs such as the i-th DAC uses the [i,i+4] bits.
The energy consumption for the uniform noise generator circuit is computed as 
\begin{equation}
    E_{noise} =\text{round}(\frac{N_{rows}}{64})E_{PNRG} + N_{rows}E_{DAC}.
\end{equation}
We consider an R2R DAC and compute its energy consumption as $E_{DAC}=I_{EO}V_{DD}t_{CLK}$ with
\begin{equation}
    I_{EO}=(V_{read}-V_{EO})\frac{1}{R_{DAC}}(1-\frac{1}{2^n_{bDAC}}).
\end{equation}

Fig.~\ref{fig:si:components}d shows the schematic of the Gaussian distributed random number generator (GPRNG) based on the Alias' method~\cite{walker1974new}.
The Alias algorithm is based on three tables implemented with Look-up Tables (LUT) in the circuit: a probability table $H$, an alias table $A$, and a non-alias table $N$.
An index $k$ for the tables is randomly generated for a PNRG.
The value stored at $H(k)$ is retrieved and compared with another random number $h$.
Based on the comparison $h>H(k)$ either the value retrieved from the alias table $A(k)$ or from the non-alias table $N(k)$ is selected.

Finally, Fig.~\ref{fig:si:components}e shows the circuit design of the Winner Takes All (WTA) circuit used for computing the $\text{argmax}$ of the chosen metric.
First, the TIA's voltage output, representing, for example, the gain values $g$, is encoded into time by a Voltage-Controlled Delay Line (VCDL).
Then, digital logic, namely a merger tree followed by arbiters and selectors,  is used to compute the variable with the highest gain by detecting the first arriving pulse from the VCDL bank.
The WTA energy is computed as
\begin{equation}
    E_{WTA}=N_{BLs}(E_{VDCL}+E_{WTA,logic})
\end{equation}
The WTA's output is connected to the XOR circuit, whose other inputs are the current state; thus, if the WTA output is high, the corresponding current state bit is flipped.
Finally, current states are stored in a register.
Registers' energy consumption is shown in Table~\ref{si:tab:params}.
Note that all the communication energy between various blocks was also simulated and modeled.

\section{Heuristics}
\begin{algorithm}[h!]
\caption{GSAT~\cite{hoos2000local}}\label{si:algo:GSAT}
\begin{algorithmic}[1]
\renewcommand{\algorithmicrequire}{\textbf{Input:}}
\renewcommand{\algorithmicensure}{\textbf{Output:}}
\Require CNF Formula $y$, Boolean input vector randomly initialized $x_0$, maximum number of iterations $\text{MAX}_{flips}$
\Ensure Input vector $x$ that maximizes the number of satisfied clauses 
\State $t \Leftarrow0$
\State $x(0) \Leftarrow x_0$
\While{$t<\text{MAX}_{flips}$ or $\text{UNSAT}$}
    \State Evaluate the number of violated clauses $w$
    \If{$w==0$}
        \State The problem is satisfied $\text{UNSAT} = False$
    \Else%[$w>0$]
        \State Compute the gain value for all the variables $g$
        \State Update the variable with the highest gain
        \State$x(t+1)\Leftarrow x(t)$
        \State $x(t+1)[\text{argmax}(g)] \Leftarrow 1 - x(t+1)[\text{argmax}(g)]$
        \State $t \Leftarrow t+1$
    \EndIf
\EndWhile
\end{algorithmic} 
\end{algorithm}

\begin{algorithm}
\caption{WalkSAT~\cite{hoos2000local}}\label{si:algo:WalkSAT}
\begin{algorithmic}[1]
\renewcommand{\algorithmicrequire}{\textbf{Input:}}
\renewcommand{\algorithmicensure}{\textbf{Output:}}
\Require CNF Formula $y$, Boolean input vector randomly initialized $x_0$, maximum number of iterations $\text{MAX}_{flips}$, random walk parameter $p$
\Ensure Input vector $x$ that maximizes the number of satisfied clauses 
\State $t \Leftarrow0$
\State $x(0) \Leftarrow x_0$
\While{$t<\text{MAX}_{flips}$ or $\text{UNSAT}$}
    \State Evaluate the number of violated clauses $w$
    \If{$w==0$}
        \State The problem is satisfied $\text{UNSAT} = False$
    \Else%[$w>0$]
        \State Select a randomly violated clause $c'$
        \State Draw a random number $p_1$
        \If{$p_0>p$}
            \State Update the member variable of $c'$ with the highest gain $v'=\textit{argmax}[g(c')]$      
            \Else
                \State Update a random member variable $v'\in c'$
            \EndIf
        \State $x(t+1)\Leftarrow x(t)$
        \State $x(t+1)[v'] \Leftarrow 1-x(t)[v']$    
        \State $t \Leftarrow t+1$
    \EndIf
\EndWhile
\end{algorithmic} 
\end{algorithm}

\begin{algorithm}
\caption{GSAT with Random Walk (GWSAT)~\cite{hoos2000local}}\label{si:algo:GWSAT}
\begin{algorithmic}[1]
\renewcommand{\algorithmicrequire}{\textbf{Input:}}
\renewcommand{\algorithmicensure}{\textbf{Output:}}
\Require CNF Formula $y$, Boolean input vector randomly initialized $x_0$, maximum number of iterations $\text{MAX}_{flips}$, random walk parameters $p$ and $wp$
\Ensure Input vector $x$ that maximizes the number of satisfied clauses 
\State $t \Leftarrow0$
\State $x(0) \Leftarrow x_0$
\While{$t<\text{MAX}_{flips}$ or $\text{UNSAT}$}
    \State Evaluate the number of violated clauses $w$
    \If{$w==0$}
        \State The problem is satisfied $\text{UNSAT} = False$
    \Else%[$w>0$]
        \State Draw a random number $p_0$
        \If{$p>p_0$}
            \State Compute the gain value for all the variables $g$
            \State Update the variable with the highest gain $v'=\textit{argmax}(g)$ 
        \Else
            \State Select a randomly violated clause $c'$
            \State Draw a random number $p_1$
            \If{$p_0>p$}
                \State Update the member variable of $c'$ with the highest gain $v'=\textit{argmax}[g(c')]$
            \Else
                \State Update a random member variable $v'\in c'$
            \EndIf
        \EndIf
        \State $x(t+1)\Leftarrow x(t)$
        \State $x(t+1)[v'] \Leftarrow 1-x(t)[v']$    
        \State $t \Leftarrow t+1$
    \EndIf
\EndWhile
\end{algorithmic} 
\end{algorithm}

\begin{algorithm}
\caption{MNSAT~\cite{pedretti2023zeroth}}\label{si:algo:MNSAT}
\begin{algorithmic}[1]
\renewcommand{\algorithmicrequire}{\textbf{Input:}}
\renewcommand{\algorithmicensure}{\textbf{Output:}}
\Require CNF Formula $y$, Boolean input vector randomly initialized $x_0$, maximum number of iterations $\text{MAX}_{flips}$, noise distribution $\mathcal{N}$, standard deviation of noise $\sigma_N$
\Ensure Input vector $x$ that maximizes the number of satisfied clauses 
\State $t \Leftarrow0$
\State $x(0) \Leftarrow x_0$
\While{$t<\text{MAX}_{flips}$ or $\text{UNSAT}$}
    \State Evaluate the number of violated clauses $w$
    \If{$w==0$}
        \State The problem is satisfied $\text{UNSAT} = False$
    \Else%[$w>0$]
        \State Compute the make value for all the variables $m$
        \State Add random noise $m' \Leftarrow \mathcal{N}(m,\sigma_N)$
        \State$x(t+1)\Leftarrow x(t)$
        \State $x(t+1)[\text{argmax}(m')] \Leftarrow 1 - x(t+1)[\text{argmax}(m')]$
        \State $t \Leftarrow t+1$        
    \EndIf
\EndWhile
\end{algorithmic} 
\end{algorithm}

\clearpage

\end{document}